# Matching Conditions in Effective-Mass Theory

Walter A. Harrison
Applied Physics Department
Stanford University
Stanford, CA 94305

It has been shown that the traditional matching of wavefunctions between regions of different effective mass (matching $\psi$ and $(1/m^*)\partial\psi/\partial x$) is not correct, but that one should match $(1/\sqrt{m^*})\psi$ and $(1/\sqrt{m^*})\partial\psi/\partial x$. It has not been clear how serious is the error in using the traditional formula. We apply the two sets of conditions to a simple, but rather general, example and find that the traditional matching is not even qualitatively correct.

Many years ago[1] we sought to calculate tunneling rates in systems describable in effective mass theory. It is familiar from basic quantum theory (e. g., Refs. 2 and 3) that at an interface when we match real wavefunctions from the two sides we must take the wavefunction to be continuous, or there will be a divergent current, and the slope of the wavefunction continuous, or there will be a divergent kinetic energy. We realized that this could not be true in effective-mass theory if the effective masses were different since then the current density $-(ie\hbar/m^*)\psi^*\partial\psi/\partial x$ would not be continuous. We could use continuity of the current as one matching condition, but were uncertain of how to choose the second. We chose to require also continuity of $\psi$ as the second condition (which turned out to be incorrect) so that our two conditions for matching between regions 1 and 2 became

$$\psi_1 = \psi_2$$
$$(m/m_1)\partial\psi_1/\partial x = (m/m_2)\partial\psi_2/\partial x. \qquad (1)$$

These conditions seem to have been almost universally used since that time (e. g., Ref. 3, p. 282).

Some years later we realized that there *was* a way to learn the second condition[4]. We could treat the simple case of a tight-binding chain which yields bands which can be described by effective mass theory at low energies, and construct the tight-binding wavefunction through the junction between regions of two different effective masses. We could then see what matching conditions would have given the correct result for this simple case. We may readily summarize how this argument went.

For a chain of atoms, spaced by $d$ and numbered $n$, each with an orbital of energy $\varepsilon_n$, and coupled to each of its neighboring orbitals by $V_{n,n-1}$ and $V_{n+1,n}$, the tight-binding equations which determine a state of energy $\varepsilon$ are[2]

$$\varepsilon_n u_n + V_{n,n-1,n} u_{n-1} + V_{n+1,n} u_{n+1} = \varepsilon u_n . \qquad (2)$$

If all $\varepsilon_n = \varepsilon_i$ and $V_{n+1,n} = V_i$ are the same, there are solutions

$$u_n = x_n + iy_n = \exp(\pm ikdn) \tag{3}$$

with energy

$$\varepsilon = \varepsilon_i + 2V_i \cos(kd). \tag{4}$$

With $V_i < 0$ these bands have a minimum at $k = 0$ and can be expanded around that minimum as $\hbar^2 k^2/2m^*$ with

$$\frac{m^*}{m} = -\frac{\hbar^2}{2md^2 V_i}. \tag{5}$$

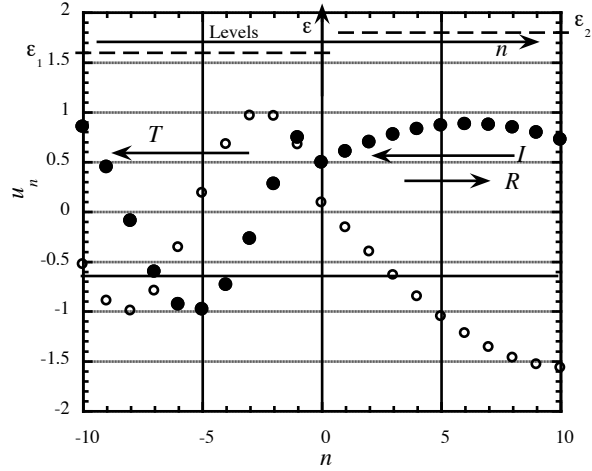

Fig. 1. States for a row of levels, of lower energy to the left and higher energy to the right, as indicated above, with different coupling on the two sides so the effective masses are different. Below are a sample set of calculated $u_n = x_n + iy_n$, with the $x_n$ as dark circles and the $y_n$ as empty circles. For negative $n$ it is a transmitted wave to the left; for positive $n$ it is incoming and reflected waves. For this case the transmission was 0.695.

We can readily solve Eq. (2) for $u_{n+1}$ in terms of $u_n$ and $u_{n-1}$. Then we construct a chain of the orbital energies and couplings of $\varepsilon_1$ and $V_1$ for negative $n$ and $\varepsilon_2$ and $V_2$ for $n \geq 0$, as indicated above in Fig. 1. For illustrative purposes we may take $V_1 = -1$ eV, which would correspond to an effective mass of $m^*/m = 1$ if the spacing were $d = 1.95$ Å, which we use on both sides. We take $V_2 = -4.$ eV, so the effective mass $m_2$ on the right is $m^*/m = 0.25$. We let $\varepsilon_2 - \varepsilon_1 = 2(V_1 - V_2) + 0.2$ eV so that the conduction-band minimum on the right is 0.2 eV above that on the left, and let the energy of an electron be 0.1 eV above the conduction-band minimum on the right, corresponding to $k_2 d = 0.158$ and $k_1 d = 0.548$. We constructed a transmitted wave $\exp(-ink_1 d)$ on the left and use the solution of Eq. (2) for $u_{n+1}$ to work through the interface at $n=0$ giving the $u_n$ shown in Fig. 1. This solution on the right is a combination of incident and reflected waves, $I\exp(-ik_2 dn) +$ $R\exp(ik_2dn)$ with $k_2$ determined from Eq. (4) for this energy. It requires only two adjacent $u_n$ to determine the values of $I$ and $R$. We may solve for the transmission of the interface at this energy as

$$Trans. = \left(1 - \frac{R^*R}{I^*I}\right) = \frac{4\sin(k_2d)(x_{n+1}y_n - y_{n+1}x_n)}{(x_{n+1} - x_n\cos(k_2d) + y_n\sin(k_2d))^2 + (y_{n+1} - y_n\cos(k_2d) - x_n\sin(k_2d))^2}$$
(6)

The transmission is the same for an electron approaching from the right at the same energy. [A similar formula was given as Eq. (8.13) in Ref. 2, but depended upon the choice of parameters on the $n<0$ side and was less general.] This provides a very powerful way of modeling complicated interfaces and tunneling systems (illustrated in Ref. 2). Note that we can make any choice of $\varepsilon_n$ and $V_n$ in the interface region and work our way through it with Eq. (2).

In Ref. 4 we treated a simple interface as above in Fig. 1, constructed states $\psi_i = A_i\cos(k_idn + \delta_i)$ on the two sides and found the matching conditions from the tight-binding Eqs. (2). It was a surprise at first to find that the most general matching conditions were required,

$$\psi_2 = A\psi_1 + B\partial\psi_1/\partial x,$$
and (7)
$$\partial\psi_2/\partial x = C\psi_1 + D\partial\psi_1/\partial x,$$

with all four coefficients nonzero. However, we noted that there had been an ambiguity in choosing the coupling between $n = -1$ and $0$, *between* the two chains. We found that the complicating terms in $B$ and $C$ disappeared *only* if we chose $V_{0,-1} = \sqrt{(V_1V_2)}$, a very plausible choice, which we then made, and which was used in calculating the points in Fig. 1. Then the matching conditions became

$$\sqrt{m/m_2}\psi_2 = \sqrt{m/m_1}\psi_1,$$
and (8)
$$\sqrt{m/m_2}\partial\psi_2/\partial x = \sqrt{m/m_1}\partial\psi_1/\partial x,$$

the only correct simple matching conditions, with the $m_i$ the $m^*$ given by Eq. (5) for each side. These conditions also, of course, conserve current. The only alternative would seem to be the full Eqs. (7). We also noted how the effective-mass equations need to be written to include spatially varying effective masses. We went on[5] to study the matching conditions if there were also second-neighbor interactions and if there were more than one orbital on each atom, which led sometimes to additional evanescent waves required in the matching.

It was never clear how serious the error was if we used the incorrect matching conditions, Eq. (1), rather than Eq. (8). We explore that here, considering again a simple interface as shown above in Fig. 1. We can directly calculate the transmission of plane waves at the interface using the traditional matching conditions from Eq. (1) and using the correct matching conditions from Eq. (8). For Eq. (8) the conditions for matching at $x = 0$, are $T/\sqrt{m_1} = (I + R)/\sqrt{m_2}$ and $-(k_1/\sqrt{m_1})T = (k_2/\sqrt{m_2})(-I + R)$, leading to a transmission of

$$Trans. = 1 - \frac{R*R}{I*I} = \frac{4k_1 k_2}{(k_1 + k_2)^2}. \tag{9}$$

We may similarly find the transmission for the traditional matching, Eq. (1). We note that the speed of the electron on each side is $v_i = \hbar k_i/m_i$. Then the transmission is given in terms of the speeds as $Trans. = 4v_1 v_2/(v_1 + v_2)^2$. This incorrect result seemed quite plausible[1], being the same as the transmission for light between two media with different refractive indices, and therefore different speeds of light. This was of course because the correct matching conditions for light are that the amplitude (the transverse electric field) is continuous but the $\partial E/\partial x$ (proportional to the transverse magnetic field) must be discontinuous to conserve energy flux. However, that condition is not appropriate for electrons, with a frequency varying as the square of the wavenumber.

We wish to compare the two formulae for transmission for various ratios $m_1/m_2$. We select a low energy on the right and let the discontinuity between the two conduction-band minima at $x = 0$ be twice as large as this energy as in the tight-binding example above. We can evaluate the two formulae for different choices of $m_1/m_2$. ($k_1^2/m_1$ is chosen equal to $3k_2^2/m_2$ to conserve energy for this case). The two results are plotted in Fig. 2 and are strikingly different, indicating that we are very much misled by using the traditional formula. They are the same only at $m_1 = m_2$, where the question of effective masses is irrelevant, giving still a transmission less than 1. As we move to a larger or small mass on the left side, the transmission varies in the opposite direction for the two formulae. The traditional formula gives a qualitatively incorrect representation of the effects of matching between regions of differing effective mass.

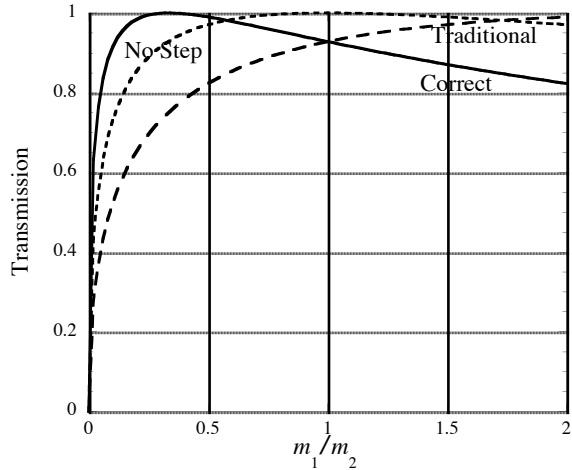

Fig. 2. The solid line is the transmission, Eq. (9), seen by an electron approaching a step in the conduction-band minimum, running off a cliff, as from the right above in Fig. 1 where we took $m_1 = 4m_2$. It is plotted as a function of the mass beyond the interface, divided by the mass as it approaches the interface. The dashed line is from the traditional formula, which could correctly represent the transmission of photons, but not electrons. Without a step in the potential, both formulae lead to the same transmission, shown as the dotted line.

We may better understand the graph by noting that in the absence of a step, both formulae lead to $\sqrt{(m_1/m_2)}/(1+\sqrt{(m_1/m_2)})^2$, independent of energy and with a peak at $m_1 = m_2$, shown as the dotted line in Fig. 2. As a step is introduced the peak for the correct formula moves toward smaller mass for the electron below the step, and that for the traditional formula toward larger mass for the electron below the step. These move symmetrically so that reversing the sign of the step and plotting against $m_2/m_1$ rather than $m_1/m_2$ gives exactly the same curves as in Fig. 2, but with the labels "Traditional" and "Correct" interchanged. Thus the $m_1/m_2 = 4$ of Fig. 1, for which the correct result was a transmission of 0.695, is the same as the traditional result in Fig. 2 for $m_1/m_2 = 0.25$. In any case, the traditional formula from Refs. 1 and 3 gives the opposite trend for the effects of effective-mass difference.